\documentclass[prd,twocolumn,nofootinbib,superscriptaddress,showpacs]{revtex4}

\usepackage{graphicx,amsmath,amsfonts,amssymb,revsymb,dcolumn,epsfig,bm}

\begin{document}
\title{Non-gaussianity in the foreground-reduced CMB maps}

\author{A. Bernui}\email{bernui@unifei.edu.br}
\affiliation{Centro Brasileiro de Pesquisas F\'{\i}sicas,
Rua Dr.\ Xavier Sigaud 150,
22290-180 Rio de Janeiro -- RJ, Brazil}
\affiliation{Instituto de Ci\^encias Exatas, Universidade Federal de Itajub\'a,
37500-903 Itajub\'a -- MG, Brazil}

\author{M.J. Rebou\c{c}as}\email{reboucas@cbpf.br}
\affiliation{Centro Brasileiro de Pesquisas F\'{\i}sicas,
Rua Dr.\ Xavier Sigaud 150,
22290-180 Rio de Janeiro -- RJ, Brazil}

\date{\today}

\begin{abstract}
A detection or nondetection of primordial non-Gaussianity by using
the cosmic microwave background Radiation (CMB) data is crucial not only to
discriminate inflationary models but also to test alternative scenarios.
Non-Gaussianity offers, therefore, a powerful probe of the physics
of the primordial universe.
The extraction of primordial non-Gaussianity is a difficult enterprise
since several effects of non-primordial nature can produce non-Gaussianity.
Given the far-reaching consequences of such a non-Gaussianity for our
understanding of the physics of the early universe, it is important to
employ a range of different statistical tools to quantify and/or
constrain its amount in order to have information that
may be helpful for identifying its causes. Moreover, different
indicators can in principle provide information about distinct
forms of non-Gaussianity that can be present in CMB data.
Most of the Gaussianity analyses of CMB data have been performed
by using part-sky frequency, where the masks are used
to deal with the galactic diffuse foreground emission.
However, full-sky  map seems to be potentially more appropriate to
test for Gaussianity of the CMB data. On the other hand, masks can
induce bias in some non-Gaussianity analyses.
Here we use two recent large-angle non-Gaussianity indicators, based on
skewness and kurtosis of large-angle patches of CMB maps,
to examine the question of non-Gaussianity in the available full-sky
five-year and seven-year Wilkinson Microwave Anisotropy Probe (WMAP) maps.
We show that these full-sky foreground-reduced maps present a
significant deviation from Gaussianity of different levels,
which vary with the foreground-reducing procedures.
We also make a Gaussianity analysis of the foreground-reduced five-year and
seven-year WMAP maps with a \emph{KQ75} mask, and compare with the similar
analysis performed with the corresponding full-sky foreground-reduced maps.
This comparison shows a significant reduction in the levels
of non-Gaussianity when the mask is employed, which provides
indications on the suitability of the foreground-reduced
maps as Gaussian reconstructions of the full-sky CMB.
\end{abstract}


\maketitle

\section{Introduction}

A key prediction of a number of simple single-field slow-roll
inflationary models is that they cannot generate detectable non-Gaussianity
of the cosmic microwave background (CMB) temperature fluctuations
within the level of accuracy of the Wilkinson Microwave Anisotropy Probe
(WMAP)~\cite{Gauss_Single-field}.
There are, however, several inflationary models that can generate  
non-Gaussianity at a level detectable by the WMAP. These non-Gaussian scenarios
comprise models based upon a wide range of mechanisms, including special
features of the inflation potential and violation of one of the following four
conditions: single field, slow roll, canonical kinetic energy, and initial
Bunch-Davies vacuum state.
Thus, although convincing detection of a fairly large primordial non-Gaussianity
in the CMB data would not rule out all inflationary models, it would exclude
the entire class of stationary models that satisfy \emph{simultaneously} these four
conditions (see, e.g., Refs.~\cite{Bartolo2004,WP-Komatsu-at-al09,Inflation-reviews}).
Moreover, a null detection of deviation from Gaussianity would rule out alternative
models of the early universe (see, for example, Refs.~\cite{NonGaus-Alternative-Models}).
Thus, a detection or nondetection of primordial non-Gaussianity in
the CMB data is crucial not only to discriminate (or even exclude classes of)
inflationary models but also to test alternative scenarios, offering therefore
a window into the physics of the primordial universe.

However, there are various non-primordial effects that can also
produce  non-Gaussianity such as, e.g., unsubtracted foreground
contamination, unconsidered point sources emission and systematic
errors~\cite{Chiang-et-al2003,Naselsky-et-al2005,Cabella-et-al2009}.
Thus, the extraction of a possible primordial non-Gaussianity is not
a  simple endeavor. In view of this, a great deal of effort
has recently gone into verifying the existence of non-Gaussianity
by employing several statistical estimators~\cite{Some_non-Gauss-refs}
(for related articles see, e.g., Refs.~\cite{Non-Gauss-related}).
Different indicators can in principle provide information about
multiple forms of non-Gaussianity that may be present in WMAP data.
It is therefore important to test CMB data for deviations from
Gaussianity by using a range of different statistical tools to
quantify or constrain the amount of any non-Gaussian signals in the
data, and extract information on their possible origins.

A  number of recent analyses of CMB data performed with different
statistical tools have provided indications of either consistency or deviation
from Gaussianity in the CMB temperature fluctuations (see, e.g.,
Ref.~\cite{Some_non-Gauss-refs}).
In a recent paper~\cite{Bernui-Reboucas2009} we proposed two new large-angle
non-Gaussianity indicators, based on skewness and kurtosis of large-angle
patches of CMB maps, which provide measures of the departure from Gaussianity
on large angular scales.
We used these indicators to search for the large-angle deviation from
Gaussianity in the three and five-year single frequency
maps with a \emph{KQ75} mask, and found that while the deviation
for the  Q, V,  and W  masked maps are within the $95\%$ expected
values of Monte-Carlo (MC) statistically Gaussian CMB maps, there is a strong
indication of deviation from Gaussianity ($\gg 95\% $ off the MC) in
the K and Ka masked maps.

Most of the Gaussianity analyses with WMAP data have been carried out by using
CMB temperature fluctuation maps (raw and clean) in the frequency bands  Q, V
and W or some combination of these maps. In these analyses, in order to deal
with the diffuse galactic foreground emission, masks such as, for example,
\emph{KQ75} and \emph{Kp0} have been used.

However, sky cuts themselves can potentially induce
bias in Gaussianity analyses, and on the other hand full-sky maps
seem more appropriate to test for Gaussianity in the CMB data. Thus,
a pertinent question that arises is how the
analysis of Gaussianity made in Ref.~\cite{Bernui-Reboucas2009}
is modified if whole-sky foreground-reduced CMB maps are used.
Our primary objective in this paper is to address this question  by extending
the analysis of Ref.~\cite{Bernui-Reboucas2009} in three different ways.
First, we use the same statistical indicators to carry out a new analysis
of Gaussianity of the available \emph{full-sky foreground-reduced} five-year and
seven-year CMB maps~\cite{ILC-5yr-Hishaw,HILC-Kim,NILC-Delabrouille,ILC-7yr-Gold}.
Second, since in these maps the foreground is reduced through different
procedures each of the resulting maps should be tested for Gaussianity.
Thus, we make a quantitative analysis of the effects of distinct cleaning
processes in the deviation from Gaussianity, quantifying  the level of
non-Gaussianity for each foreground reduction method.
Third, we study quantitatively the consequences for the Gaussianity analysis
of masking the foreground-reduced maps with the \emph{KQ75} mask.
An interesting outcome is that this mask lowers significantly the level of
deviation from Gaussianity even in the foreground-reduced maps, rendering
therefore information about the suitability of the foreground-reduced maps
as Gaussian reconstructions of the full-sky CMB.

\section{Non-Gaussianity Indicators} \label{Sec:Indicators}

The chief idea behind our construction of the non-Gaussianity indicators is that
a simple way of accessing the deviation from Gaussianity distribution of the
CMB temperature fluctuations  is by calculating  the skewness $S=\mu_3/\sigma^3$,
and the kurtosis $K=\mu_4/\sigma^4 - 3$ from the fluctuations data,
where $\mu_3$ and $\mu_4$ are the third and fourth central moments of the
distribution, and $\sigma$ is its variance.
Clearly calculating $S$ and $K$ from  the whole sky
temperature fluctuations data would simply yield two dimensionless numbers,
which are rough measures of deviation from Gaussianity of the temperature
fluctuation distribution.

However, one can go further and obtain a great number of values
associated to directional information of deviation from Gaussianity
if instead one takes a discrete set of points $\{j=1, \ldots ,N_\mathrm{c} \}$
homogeneously distributed on the celestial sphere $S^2$ as the center of
spherical caps of a given aperture $\gamma$ and calculate $S_j$ and $K_j$
from the CMB temperature fluctuations of each spherical cap.
The values $S_j$ and $K_j$ can then be taken as measures of the
non-Gaussianity in the direction $(\theta_j, \phi_j)$ of the center
of the spherical cap $j\,$. Such calculations for the individual
caps thus provide quantitative information ($2 N_\text{c}$ values) about
possible violation of Gaussianity in the CMB data.

This procedure is a constructive way of defining two discrete functions
$S$ and $K$ (defined on $S^2)$ from the temperature fluctuations
data, and  can be formalized through the following steps (for more
details, see Ref.~\cite{Bernui-Reboucas2009}):
\begin{enumerate}
\item[{\bf i.}]
Take a discrete set of points $\{j=1, \ldots ,N_{\rm c}\}$ homogeneously
distributed on the CMB celestial sphere $S^2$ as the centers of spherical
caps of a given aperture $\gamma$;
\item[{\bf ii.}]
Calculate for each spherical cap $j$ the skewness ($S_j$) and kurtosis ($K_j$)
given, respectively,  by
\begin{eqnarray}
&&S_j   =  \frac{1}{N_{\rm p} \,\sigma^3_{\!j} } \sum_{i=1}^{N_{\rm p}}
\left(\, T_i\, - \overline{T_j} \,\right)^3 \,, \label{S-Def}\\
\text{and} && \nonumber  \\
&&K_j   =  \frac{1}{N_{\rm p} \,\sigma^4_{\!j} } \sum_{i=1}^{N_{\rm p}}
\left(\,  T_i\, - \overline{T_j} \,\right)^4 - 3 \label{K-Def} \,,
\end{eqnarray}
where $N_{\rm p}$ is the number of pixels in the $j^{\,\rm{th}}$ cap,
$T_i$ is the temperature at the $i^{\,\rm{th}}$ pixel, $\overline{T_j}$ is
the CMB mean temperature in the $j^{\,\rm{th}}$ cap, and $\sigma$ is the
standard deviation. Clearly, the values $S_j$ and $K_j$ obtained in this
way for each cap can be viewed as a measure of non-Gaussianity in the
direction of the center of the cap $(\theta_j, \phi_j)$;
\item[{\bf iii.}]
Patching together the $S_j$ and $K_j$ values for each spherical cap,
one obtains our indicators, i.e., discrete functions $S = S(\theta,\phi)$
and $K = K(\theta,\phi)$ defined over the celestial sphere, which can be
used to measure the deviation from Gaussianity as a function of the angular
coordinates $(\theta,\phi)$. The Mollweid projection of skewness and kurtosis
functions $S = S(\theta,\phi)$ and $K = K(\theta,\phi)$  are nothing but
skewness and kurtosis maps, hereafter we shall refer to them as $S-$map
and $K-$map, respectively.
\end{enumerate}

Now, since  $S = S(\theta,\phi)$ and $K = K(\theta,\phi)$ are
functions defined on $S^2$ they can be expanded into their spherical
harmonics in order to have their power spectra $S_{\ell}$ and $K_{\ell}$.
Thus, for example, for the skewness indicator $S = S(\theta,\phi)$ one has
\begin{equation}
S (\theta,\phi) = \sum_{\ell=0}^\infty \sum_{m=-\ell}^{\ell}
b_{\ell m} \,Y_{\ell m} (\theta,\phi) \; ,
\end{equation}
and can calculate the corresponding angular power spectrum
\begin{equation}
S_{\ell} = \frac{1}{2\ell+1} \sum_m |b_{\ell m}|^2 \; ,
\end{equation}
which can be used to quantify the angular scale of
the deviation from Gaussianity, and also to calculate
the statistical significance of such deviation.
Obviously, similar expressions hold for the kurtosis $K = K(\theta,\phi)$.

In the next section we shall use the statistical indicators
$S = S(\theta,\phi)$ and $K = K(\theta,\phi)$ to test for Gaussianity
the available foreground-reduced maps obtained from the five-year WMAP
data.

\section{Non-Gaussianity}

\begin{figure*}[htb!]
\begin{center}
\includegraphics[width=5.0cm,height=8.2cm,angle=90]{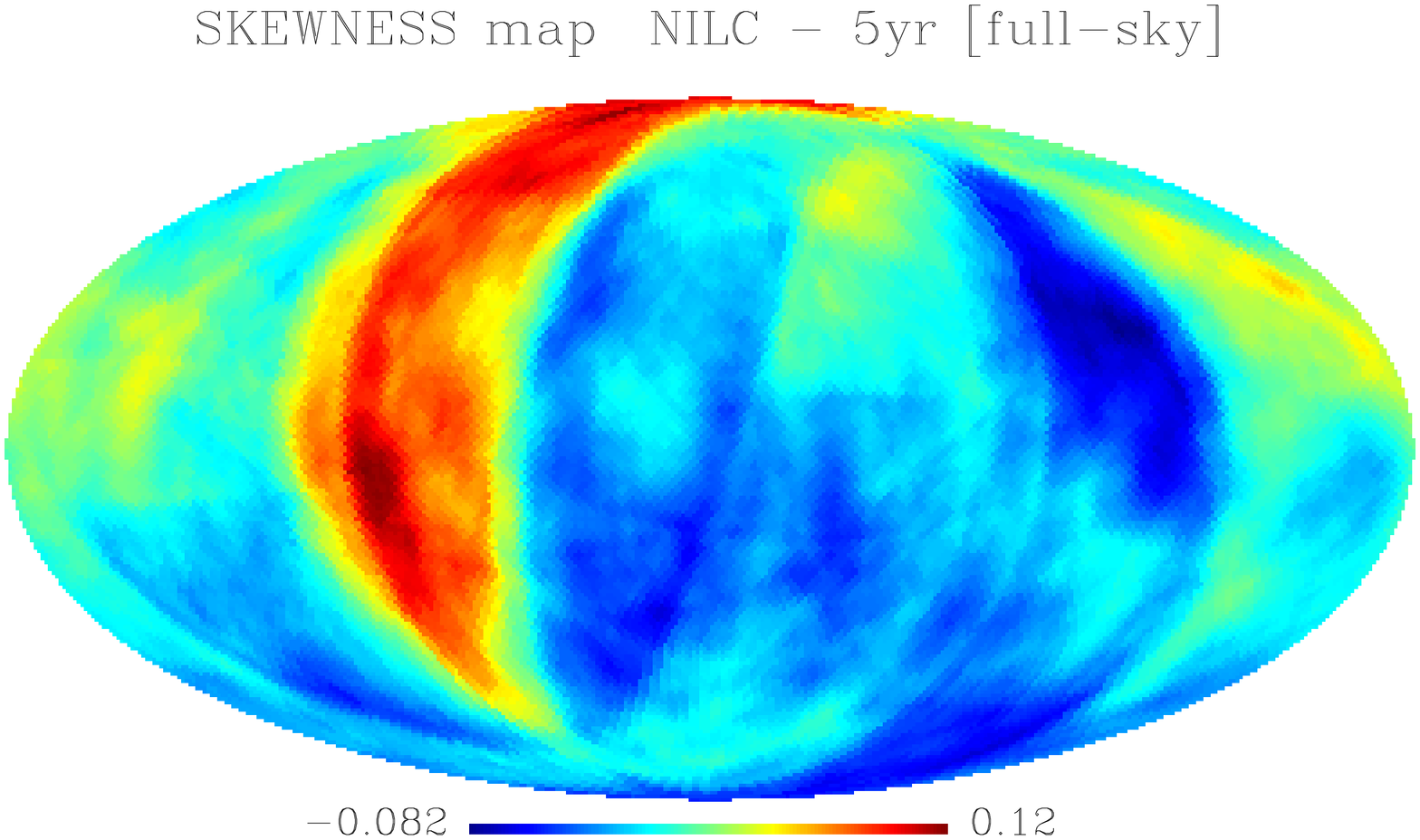}
\hspace{5mm}
\includegraphics[width=5.0cm,height=8.2cm,angle=90]{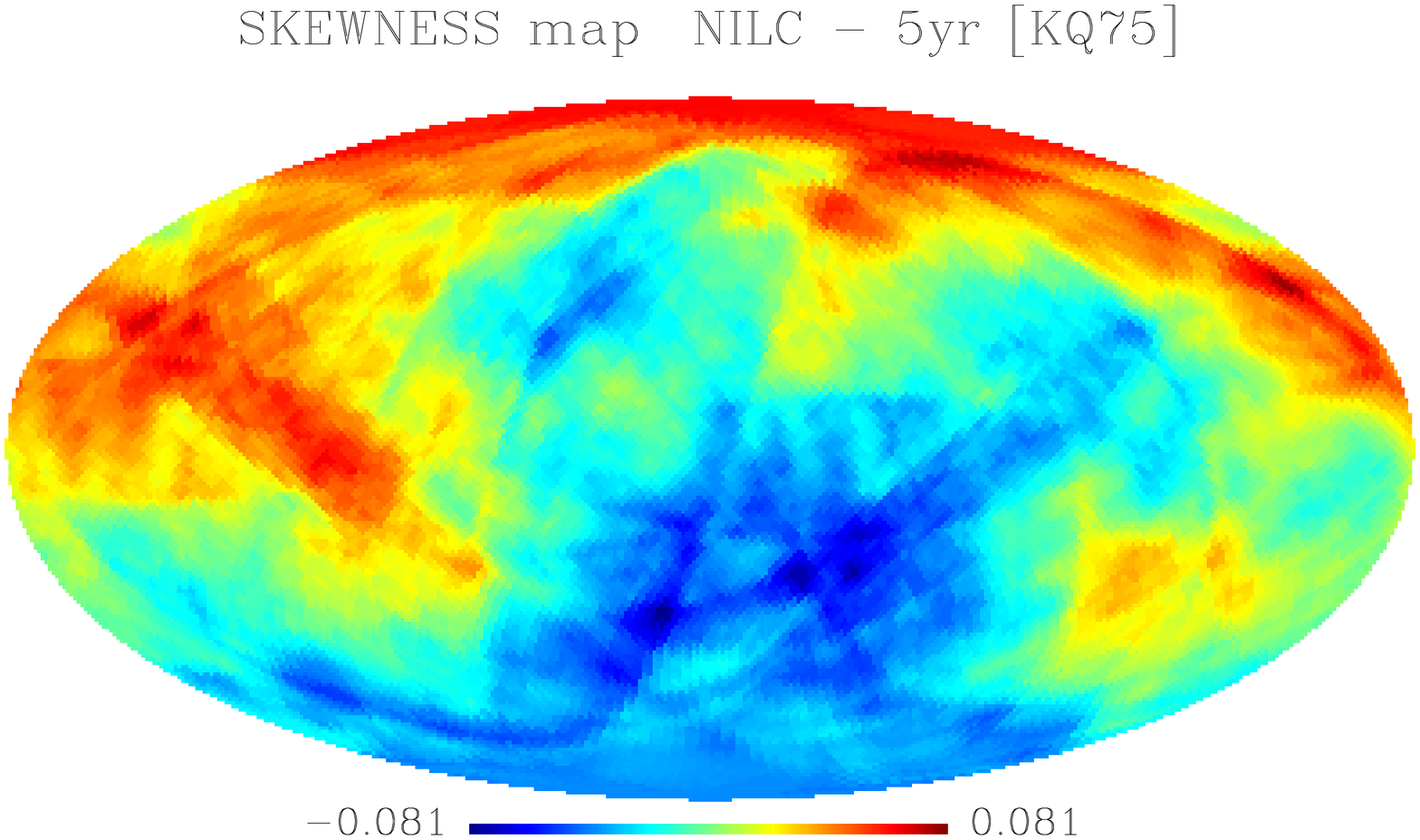}
\caption{\label{Fig1} Skewness indicator  maps calculated from
the  five-year foreground-reduced NILC full-sky (left panel) and
\emph{KQ75} masked (right panel) maps.}
\end{center}
\end{figure*}

\begin{figure*}[htb!]
\begin{center}
\includegraphics[width=5.0cm,height=8.2cm,angle=90]{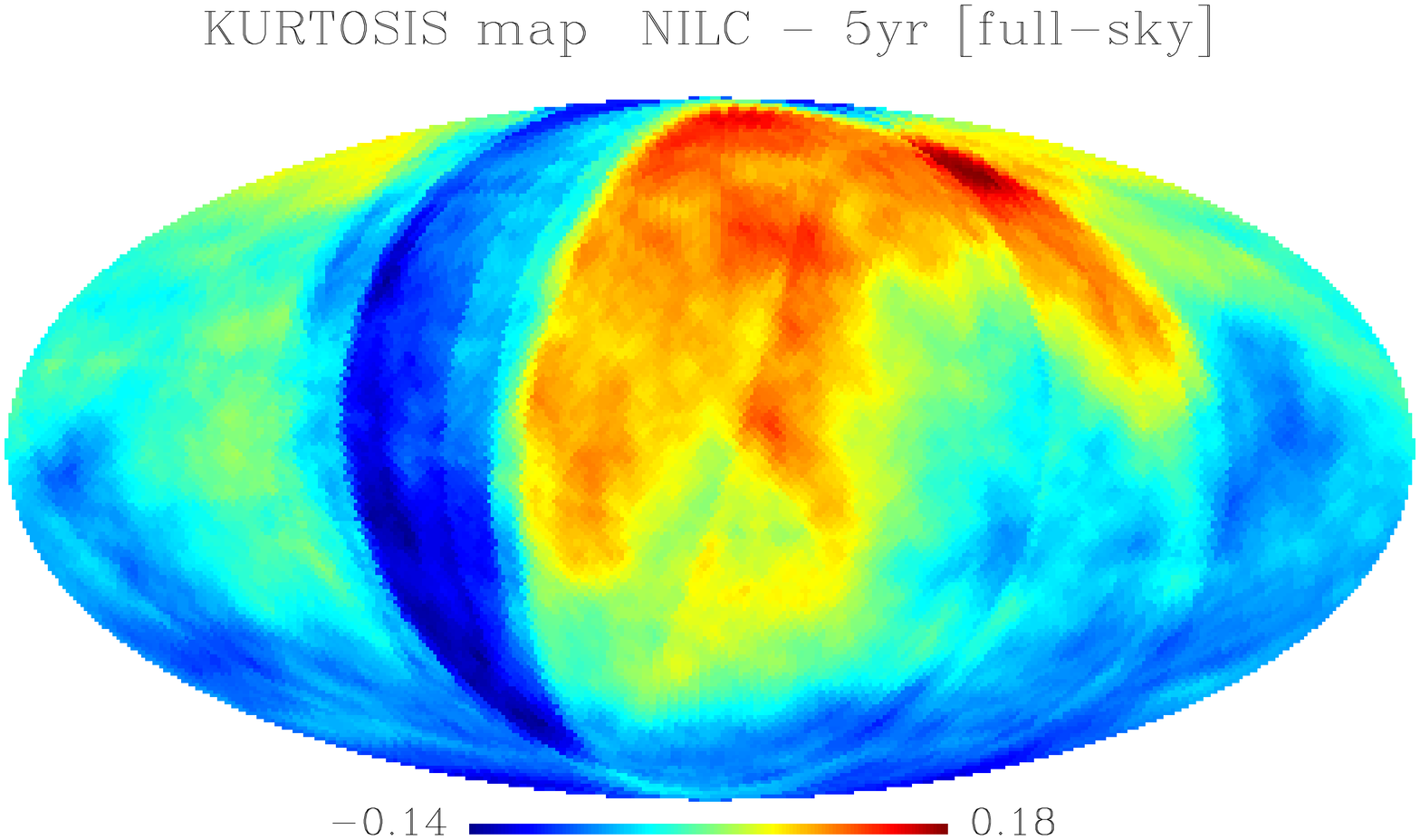}
\hspace{5mm}
\includegraphics[width=5.0cm,height=8.2cm,angle=90]{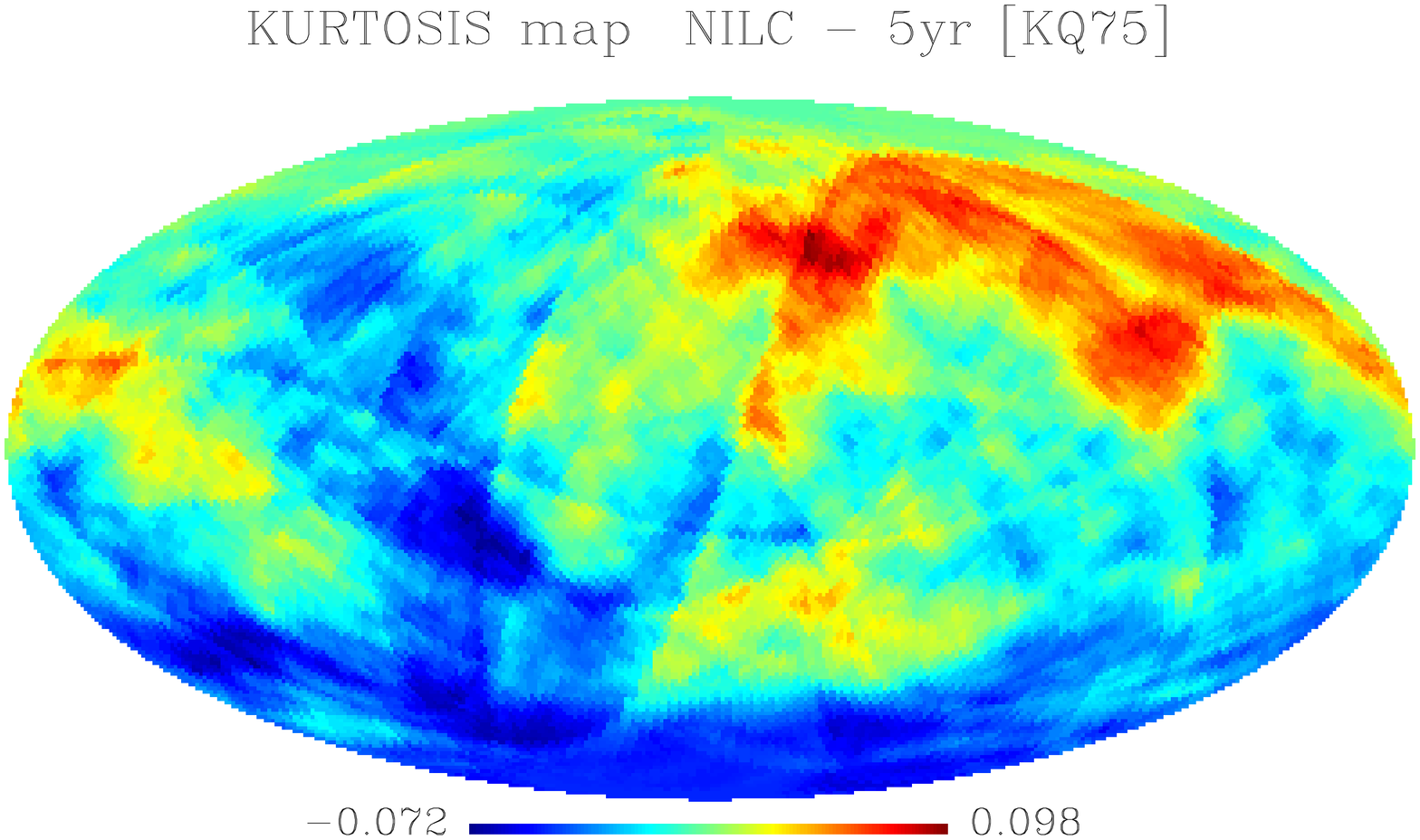}
\caption{\label{Fig2} Kurtosis indicator maps
calculated from the five-year foreground-reduced NILC
full-sky (left panel) and \emph{KQ75} masked (right panel) maps.}
\end{center}
\end{figure*}

\subsection{Foregound-reduced maps}

The WMAP team has released high angular resolution  five-year maps of
the CMB temperature fluctuations in the five frequency bands K ($22.8$ GHz),
Ka ($33.0$ GHz), Q ($40.7$ GHz), V ($60.8$ GHz), and W ($93.5$ GHz). They
have also produced a full-sky foreground-reduced Internal Linear Combination
(ILC) map which is formed from a weighted linear combination of these five
frequency band maps in which the weights are chosen in order to minimize
the galactic foreground contribution.

It is well known that the \textit{first-year} ILC map is inappropriate for CMB scientific
studies~\cite{Bennett2003}. However, in the \textit{five-year} (also in the three-year and
seven-year) version of this map a bias correction has been implemented as part of the
foreground cleaning process, and the WMAP team suggested that this map is suitable for use
in large angular scales (low $\ell$) analyses although they admittedly  have not performed
non-Gaussian tests on this version of the ILC map~\cite{ILC-3yr-Hishaw,ILC-5yr-Hishaw}.
Notwithstanding the many merits of the five-year ILC procedure, some cleaning features of
this ILC approach have been considered, and two variants have been proposed recently.
In the first approach  the frequency dependent weights were  determined in harmonic
space~\cite{HILC-Kim}, while in the second the foreground is reduced by using
needlets as the basis of the cleaning process~\cite{NILC-Delabrouille}.
Thus, two new full-sky foreground-cleaned maps have been produced with the WMAP
five-year data, namely the harmonic ILC (HILC)~\cite{HILC-Kim} and the needlet
ILC (NILC) (for more details see Refs.~\cite{HILC-Kim,NILC-Delabrouille}).

In the next section, we use the full-sky foreground-reduced ILC, HILC and NILC
maps with the same smoothed $1^{\circ}$ resolution (which is the resolution of the
ILC map) as the input maps from which we calculate the $S = S(\theta,\phi)$ and
$K = K(\theta,\phi)$ maps, and then we compute the associated power spectra in order to
carry out a statistical  analysis to quantify the levels of deviation from Gaussianity.%
\footnote{The ILC, HILC and NILC maps are available for download from:
http://lambda.gsfc.nasa.gov/product/map/dr3/ilc\_map\_get.cfm,
http://www.nbi.dk/$\sim$jkim/hilc/ and
http://www.apc.univ-paris7.fr/APC\_CS/Recherche/Adamis/cmb\_wmap-en.php.}

\subsection{Analysis and results} \label{Anal_Resul}

In order to minimize the statistical noise, in the calculations
of skewness and kurtosis maps ($S-$map and $K-$map) from the
foreground-reduced maps, we have scanned the celestial
sphere with spherical caps of aperture  $\gamma = 90^{\circ}$,
centered at $12\,288$ points homogeneously generated on the two-sphere
by using  the HEALPix code~\cite{Gorski-et-al-2005}.
In other words, the point-centers of the spherical caps are the center of
the pixels of a homogeneous pixelization  of the $S^2$
generated by HEALPix  with $N_{\text{side}}=32$.
We emphasize, however, that this pixelization is only a practical way
of choosing the centers of the caps homogeneously distributed on $S^2$.
It is not related to the pixelization of the above-mentioned ILC, HILC
and NILC input maps that we have utilized to calculate both the $S$ and $K$
maps from which we compute the associated power spectra.

Figures~\ref{Fig1} and~\ref{Fig2}  show examples  of $S$ and $K$ maps
obtained from the foreground-reduced NILC full-sky and \emph{KQ75} maps.
The panels of these figures clearly show regions with higher and lower
values ('hot' and 'cold' spots) of  $S(\theta,\phi)$ and $K(\theta,\phi)$,
which suggest \emph{large-angle}  multipole components of non-Gaussianity.
We have also calculated similar maps (with and without the \emph{KQ75} mask)
from the ILC and HILC maps. However, since these maps provide only
\emph{qualitative} information, to avoid repetition we only depict
the maps of Figs.~\ref{Fig1} and~\ref{Fig2}
merely for illustrative purpose.

In order to obtain \emph{quantitative} information about the large angular
scale (low $\ell$) distributions for the non-Gaussianity $S$ and $K$ maps
obtained from the available full-sky foreground-reduced five-year maps,
we have calculated the (low $\ell$) power spectra $S_{\ell}$ and $K_{\ell}$
for these maps. 
The statistical significance of these power spectra is estimated
by comparing with the corresponding multipole values
of the averaged power spectra $\overline{S}_{\ell}$ and $\overline{K}_{\ell}$
calculated from maps obtained by averaging over $1\,000$ Monte-Carlo-generated
statistically Gaussian CMB maps.%
\footnote{Each Monte-Carlo \emph{scramb\/led} map is a stochastic
realization of the WMAP best-fitting angular power spectrum of the
$\Lambda$CDM model, obtained by randomizing the temperature components
$a_{\ell m}$ within the cosmic variance limits.}
Throughout the paper the mean quantities are denoted by overline.

Before proceeding to a statistical analysis, let us describe
with some detail our calculations. For the sake of brevity,
we focus on the skewness indicator $S$, but a completely
similar procedure was used for the kurtosis indicator $K$.
We generated $1\,000$ MC Gaussian (\emph{scramb\/led})
CMB maps, which are then used to generate $1\,000$ skewness $S-$maps,
from which we calculate $1\,000$ power spectra:
$\{S^{\,\mathbf{i} }_{\,\ell}\}$ ($\,\mathbf{i}= 1,\,\,\cdots,1\,000\,$
is an enumeration index, and $\,\ell=1,\,\,\cdots,10\,$).
In this way, for each fixed multipole component
$S^{\,\mathbf{i} }_{\ell=\text{fixed}}$ we have $1\,000\,$ multipole values
from which we calculate the mean value $\overline{S}_{\ell} =
(1/1000) \sum_{\mathbf{i} = 1}^{1000} S_{\,\ell}^{\,\mathbf{i}}\,$.
{}From this MC process we have at the end ten mean multipole values
$\overline{S}_{\ell}$, each of which are then used for a comparison
with the corresponding multipole values $S_{\ell}$ (obtained from
the input map) in order to evaluate the statistical significance of the
multipole components $S_{\ell}$.
To make this comparison easier, instead of using the angular power spectra
$S_{\ell}$ and $K_{\ell}$ themselves, we  employed the \emph{differential} power
spectra $|S_{\ell} - \overline{S}_{\ell}|$ and $|K_{\ell} - \overline{K}_{\ell}|$,
which measure the deviation of the skewness and kurtosis multipole values
(calculated from the foreground-reduced maps) from the mean multipoles
$\overline{S}_{\ell}$ and $\overline{K}_{\ell}$ (calculated from the
Gaussian maps).
Thus, for example, to study the statistical significance of the quadrupole
component of the skewness from HILC map $S_{2}^{\,\text{\sc HILC}}$ (say) we
calculate the deviation $|S_{2}^{\,\text{HILC}} - \overline{S}_{2}|$,
where the mean quadrupole value $\overline{S}_{2}$ is calculated from the
$\,\mathbf{i}= 1,\,\,\cdots,1\,000\,$ quadrupole values of the
MC Gaussian maps.
\begin{figure*}[htb!]
\begin{center}
\includegraphics[width=8.8cm,height=5.6cm]{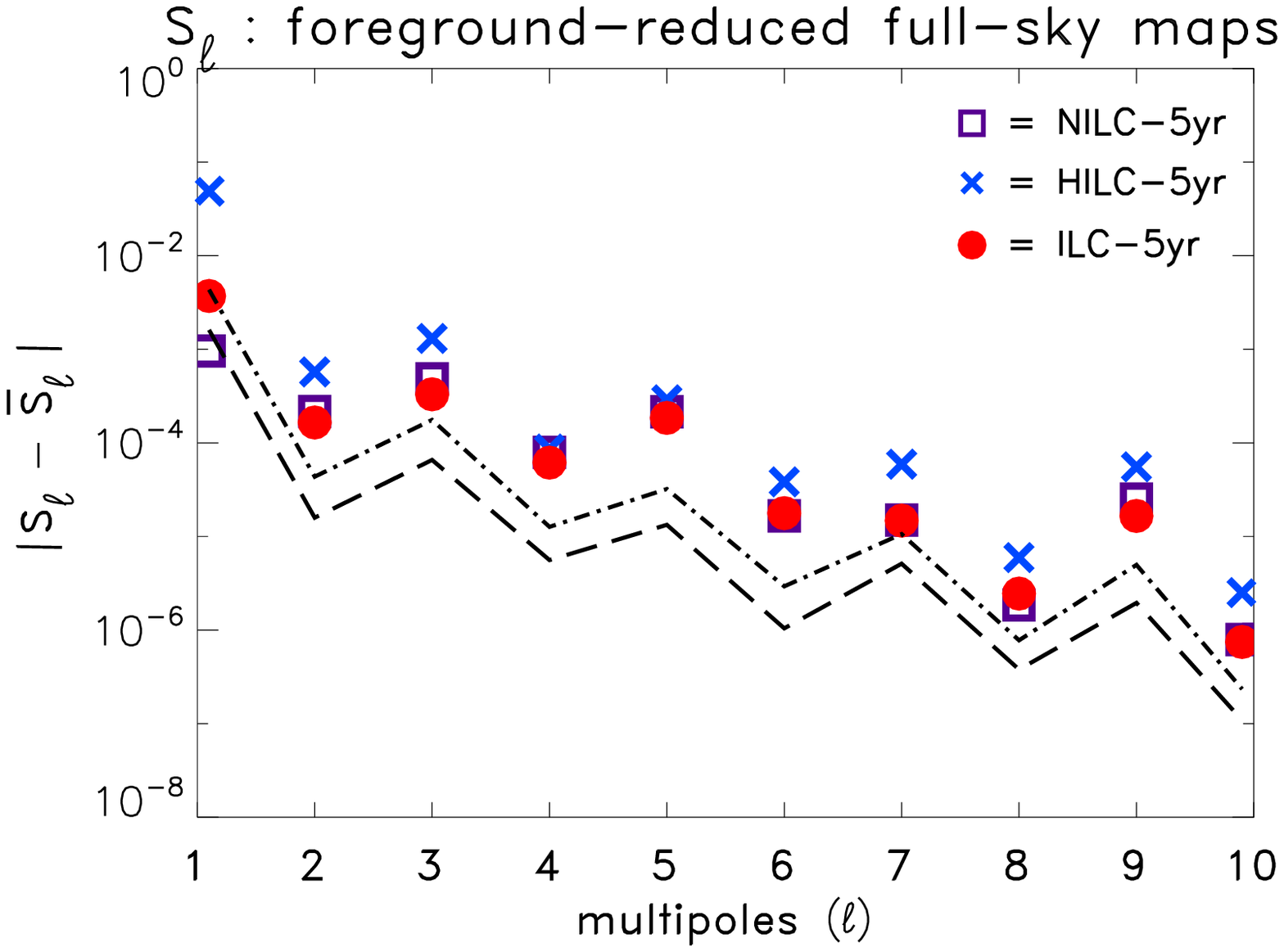}
\includegraphics[width=8.8cm,height=5.6cm]{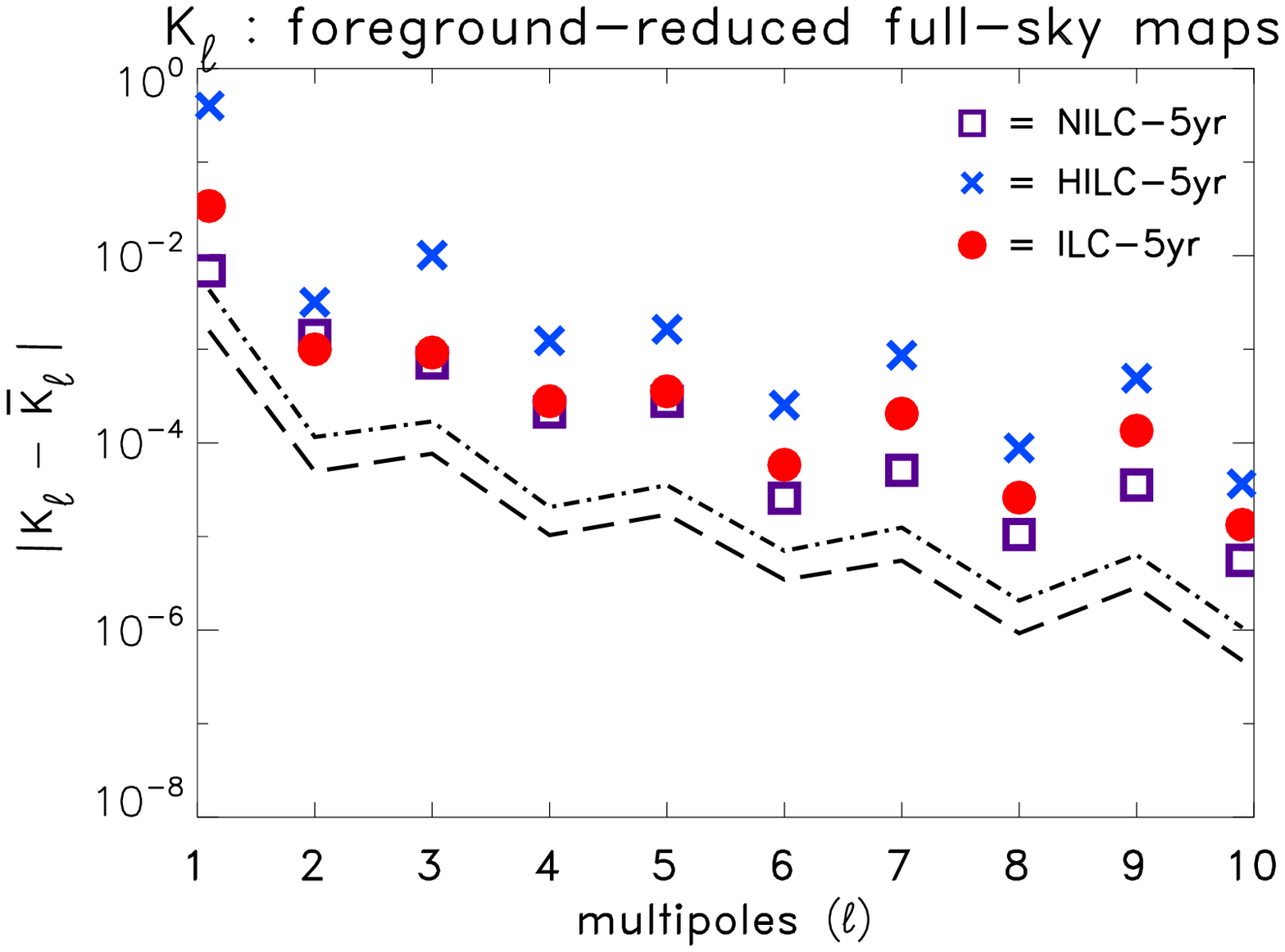}
\caption{Differential power spectrum of skewness $|S_{\ell} - \overline{S}_{\ell}|$
(left) and kurtosis $|K_{\ell} - \overline{K}_{\ell}|$ (right) indicators calculated
from the full-sky foreground-reduced ILC, HILC, and NILC maps obtained from
the WMAP five-year data. The $68\%$ and $95\%$ confidence levels are
indicated, respectively, by the dashed and dash-dotted lines.}
\label{Fig3}
\end{center}
\end{figure*}

\begin{table*}[!bht]
\begin{center}
\begin{tabular}{ccc} 
\hline \hline 
       \ \  & \ \ \ $\chi^2_{}$ for $S_\ell$ \ \ \    & \ \ \ $\chi^2_{}$ for $K_\ell$ \ \ \   \\
\hline
HILC   \ \  & \ \ \ $4\,625$ \ \ \   & \ \ \ $301\,665$ \ \ \      \\
ILC    \ \  & \ \ \ $35.7$    \ \ \  & \ \ \ $2\,368$     \ \ \     \\
NILC   \ \  & \ \ \ $ 7.1$      \ \ \ & \ \ \ $160.3$   \ \ \     \\
\hline \hline
\end{tabular}
\end{center}
\caption{Results of the $\chi^2$ test  to determine 
the goodness of fit for $S_{\ell}$ and $K_{\ell}$ multipole values
calculated from the full-sky foreground-reduced HILC, ILC, and
NILC maps as compared to the expected multipoles values from
the Gaussian MC maps.}
\label{Chi2-Table}
\end{table*}

Figure~\ref{Fig3} shows the differential power spectra calculated from
\emph{full-sky} five-year foreground-reduced maps, i.e., it displays the
absolute value of the deviations from the mean angular power spectrum
of the skewness $S_{\ell}$ (left panel) and kurtosis $K_{\ell}$ (right panel)
indicators for $\,\ell=1,\,\,\cdots,10\,$, which is a range of multipole
values needed to investigate the large-scale angular characteristics of the
$S$ and $K$ maps.
This figure shows a first indication of deviation
from Gaussianity in five-year foreground-reduced ILC, HILC and NILC maps
in that the deviations $|S_{\ell} - \overline{S}_{\ell}|$ and
$|K_{\ell} - \overline{K}_{\ell}|$ for these maps are not
within $95\%$ of the mean MC value.

To obtain additional quantitative information regarding the deviation
from Gaussianity, we can also calculate the percentage of the deviations
$|S^{\,\mathbf{i} }_{\,\ell} - \overline{S}_{\ell}|$ calculated from
$1\,000$  MC Gaussian maps,  which are smaller than
$|S_{\ell} - \overline{S}_{\ell}|$ obtained from each foreground-reduced
map. This calculations are made in detail in the Appendix~\ref{App_A}.
Thus, for example,
we have for the full-sky NILC, HILC and ILC maps, respectively,
that $\sim 99.999\%$, $\sim 99.999\%$, and $99.900\%$ of the multipole values
$S^{\,\mathbf{i} }_{5}$ obtained from the  MC maps are closer to the
mean $\overline{S}_{5}$ than the value $S_{5}$ calculated from the data,
i.e. from each of the foreground-reduced maps.
This indicates how unlikely are the occurrences of the values obtained
from these foreground-reduced maps for the multipole $S_{5}$ in
the set of values of $S^{\,\mathbf{i} }_{5}$ from MC simulated maps.
In other words, the probability of occurrence of the $S_{5}$ values
(in the set of MC values) for the NILC, HILC and ILC maps is only
$\mathcal{O} (10^{-3})\%$, $\mathcal{O} (10^{-3})\%$ and $\mathcal{O} (10^{-1})\%$,
respectively. Similarly, the probability of occurrence of $K_2$, for example,
is $\mathcal{O} (10^{-3})\%$ for all these foreground-reduced maps,
while for $K_5$ are respectively $\mathcal{O} (10^{-1})\%$ (NILC),
$\mathcal{O} (10^{-3})\%$ (HILC) and $\mathcal{O} (10^{-3})\%$ (ILC).
In Tables~\ref{Skew-dev_prob_full} and~\ref{Kurt-dev_prob_full} of the Appendix~\ref{App_A}
we collect together the probability of occurrence of each of the values
$S_{\ell}$ and $K_{\ell}$ ($\,\ell=1,\,\cdots,10\,$) calculated from $S$ and $K$ maps
obtained from the full-sky NILC, HILC and ILC maps. In Tables~\ref{Skew-dev_prob_masked},
and~\ref{Kurt-dev_prob_masked} we present these probabilities
calculated from the same input maps but now with \emph{KQ75} mask.%
\footnote{We emphasize that, throughout this paper, in the implementation of the mask
we do not take $T=0$ for the temperature fluctuation of the pixels inside the masked
region. This would clearly induce non-Gaussian contribution. In our scan of the CMB sky
when the spherical cap move into the masked area the pixels of the cap inside the
masked do not contribute to the values of the indicators in the center of the cap.
In these cases, the values $S_j$ and $K_j$ for a  $j^{\,\rm{th}}$ cap
are calculated with  small number $N_{\rm p}$ of pixels.}
The comparison of Table~\ref{Skew-dev_prob_full} with
Table~\ref{Skew-dev_prob_masked}, and of Table~\ref{Kurt-dev_prob_full}
with Table~\ref{Kurt-dev_prob_masked}, makes apparent the role of the \emph{KQ75}
mask in reducing the level of deviation from Gaussianity (see the Appendix~\ref{App_A} for more details).

\begin{figure*}[htb!]
\begin{center}
\includegraphics[width=8.8cm,height=5.6cm]{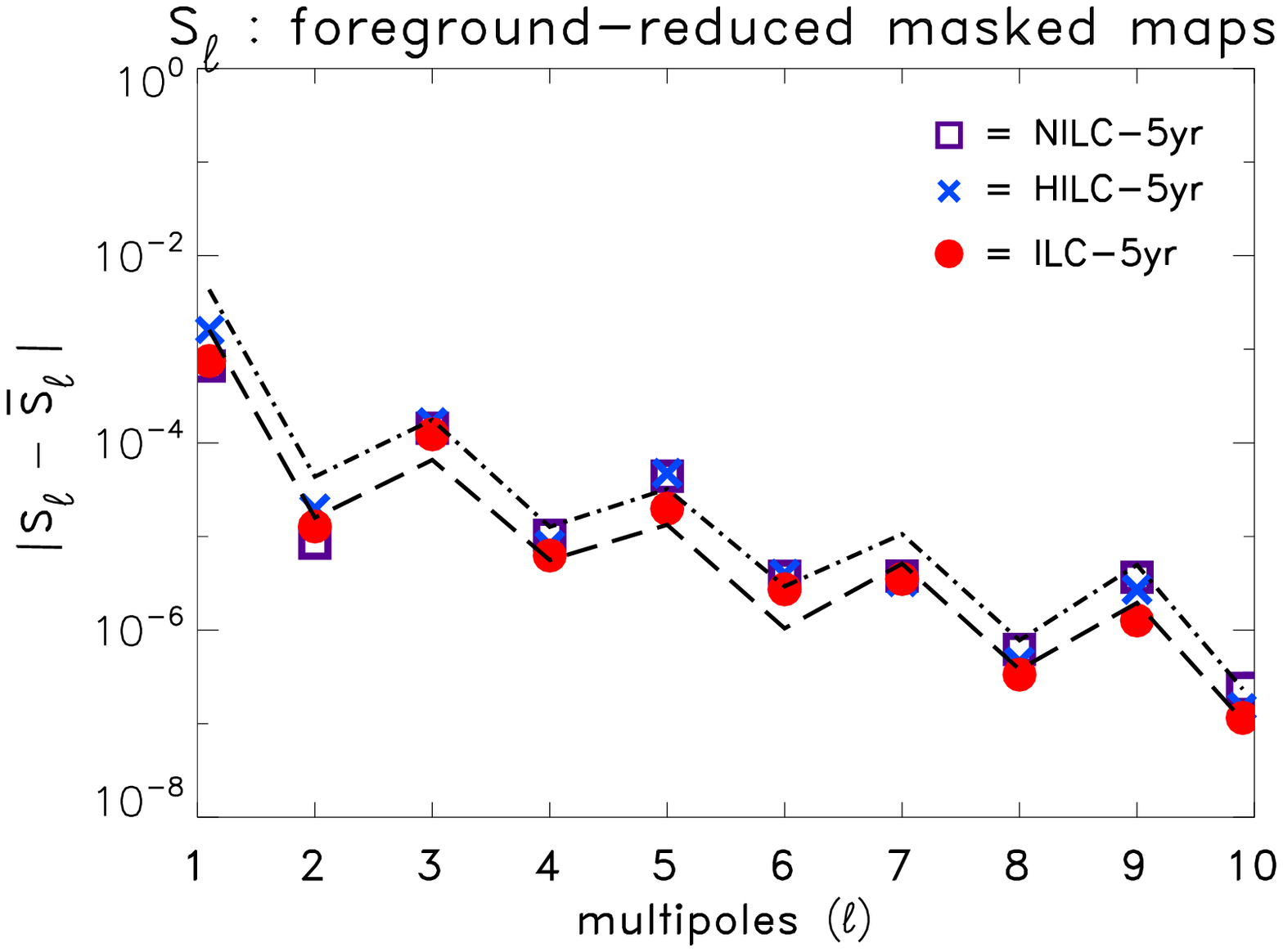}
\includegraphics[width=8.8cm,height=5.6cm]{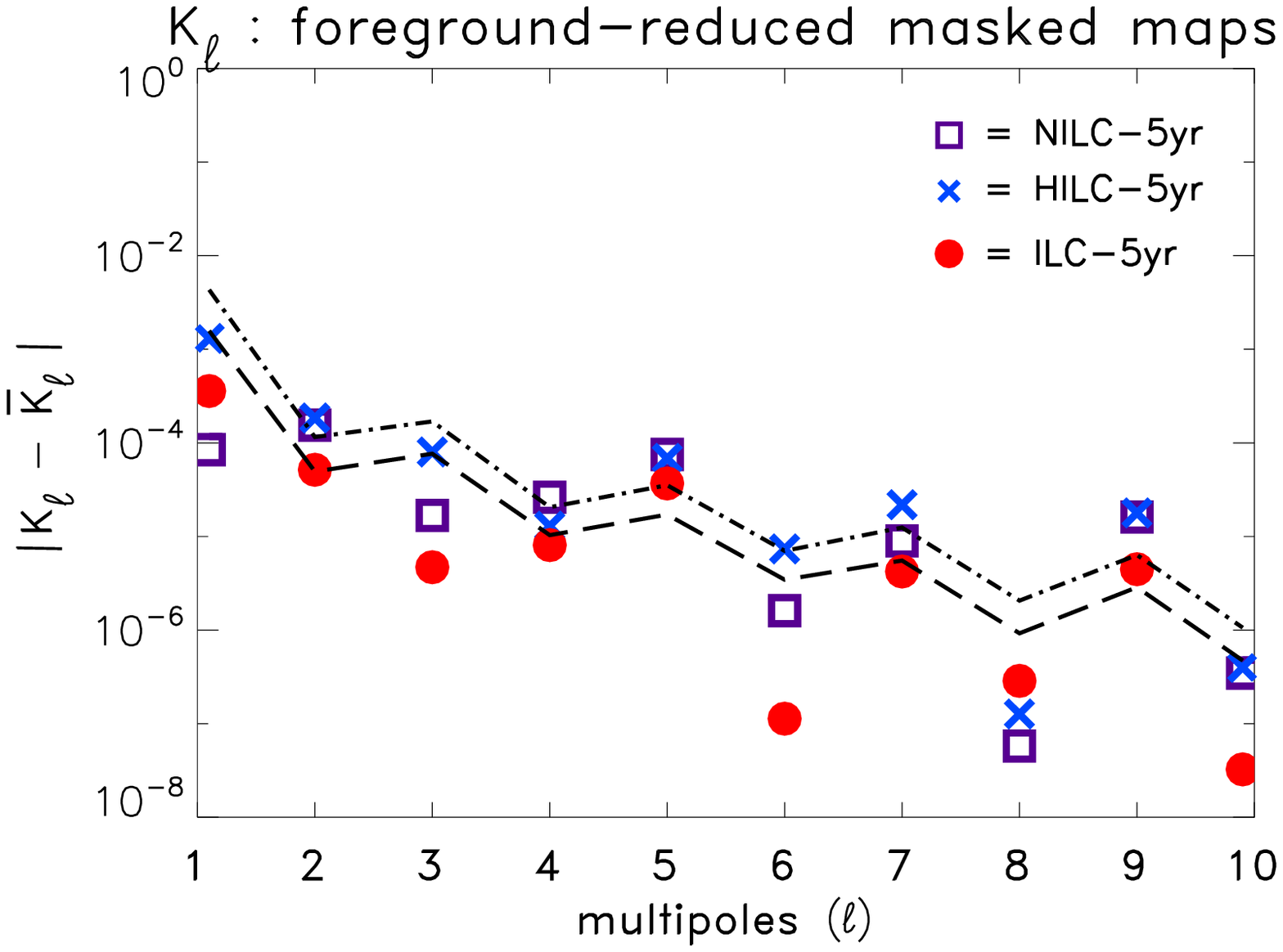}
\caption{\label{Fig4} Differential power spectrum of skewness
$|S_{\ell} - \overline{S}_{\ell}|$ (left) and kurtosis
$|K_{\ell} - \overline{K}_{\ell}|$ (right) indicators calculated
from the five-year foreground-reduced \emph{KQ75} masked ILC, HILC
and NILC maps. The $68\%$ and $95\%$ confidence levels are
indicated, respectively, by the dashed and dash-dotted lines.}
\end{center}
\end{figure*}

\begin{table*}[!bht]
\begin{center}
\begin{tabular}{ccc} 
\hline \hline 
\ \  & \ \ \ $\chi^2_{}$ for $S_\ell$ [\emph{KQ75}] \ \ \
& \ \ \ $\chi^2_{}$ for $K_\ell$ [\emph{KQ75}]
\ \ \      \\
\hline
HILC   \ \  & \ \ \ $4.7$ \ \ \   & \ \ \ $4.2$ \ \ \      \\
ILC    \ \  & \ \ \ $1.2$    \ \ \  & \ \ \ $0.4$     \ \ \     \\
NILC   \ \  & \ \ \ $1.4$      \ \ \ & \ \ \ $1.1$   \ \ \     \\
\hline \hline
\end{tabular}
\end{center}
\caption{Results of the $\chi^2$ test  to determine
the goodness of fit for $S_{\ell}$ and $K_{\ell}$ multipole values
calculated from the foreground-reduced HILC, ILC, and
NILC maps with a \textit{KQ75} mask as compared to the expected
multipoles values from the Gaussian MC masked maps.}
\label{Chi2-Table-masked}
\end{table*}

\begin{table*}[!bht]
\begin{center}
\begin{tabular}{ccc} 
\hline \hline 
\ \  & \ \ \ $\chi^2_{}$ for $S_\ell$ [\emph{Kp0}] \ \ \ & \ \ \ $\chi^2_{}$ for $K_\ell$ [\emph{Kp0}]
\ \ \      \\
\hline
HILC   \ \  & \ \ \ $58.7$ \ \ \   & \ \ \ $101.9$ \ \ \      \\
ILC    \ \  & \ \ \ $1.9$    \ \ \  & \ \ \ $6.5$     \ \ \     \\
NILC   \ \  & \ \ \ $4.5$      \ \ \ & \ \ \ $17.9$   \ \ \     \\
\hline \hline
\end{tabular}
\end{center}
\caption{Results of the $\chi^2$ test  to determine
the goodness of fit for $S_{\ell}$ and $K_{\ell}$ multipole values
calculated from the $S$ and $K$ maps obtained from the foreground-reduced HILC, ILC, and
NILC maps with a \textit{Kp0} mask as compared to the expected
multipoles values from the Gaussian MC masked maps.}
\label{Chi2-Table-kp0mask}
\end{table*}

Although the set of  'local' (fixed $\ell$) estimates collected together
in the tables of Appendix~\ref{App_A} gives an indication
of deviation from Gaussianity as measured by each multipole component
to have an overall assessment of low $\ell$ power spectra $S_\ell$ and $K_\ell$
calculated from each CMB foreground-reduced map, we have performed a $\chi^2$
test to find out the goodness of fit for $S_{\ell}$ and $K_{\ell}$ multipole
values as compared to the expected multipole values from the MC Gaussian maps.
In this way, we can obtain one number for each foreground-reduced map that
collectively ('globally') quantifies the deviation from Gaussianity.
For the power spectra $S_\ell$ and  $K_\ell$ we found that the values
given in Table~\ref{Chi2-Table} for the ratio $\chi^2/\text{dof}\,$
(dof stands for degrees of freedom) for the power spectra calculated
from HILC, ILC and  NILC full-sky input maps.
Clearly a good fit occurs when $\chi^2/\text{dof}\,\sim 1 $. Moreover,
greater are the $\chi^2/\text{dof}\,$  values, the smaller
the $\chi^2$ probabilities, that is the probability that the multipole
values $S_{\ell}$ and $K_{\ell}$ and the expected MC multipole values agree.
Thus, regarding the skewness indicator Table~\ref{Chi2-Table} shows
that the HILC presents the greatest level of deviation from Gaussianity
($\chi^2/\text{dof}\,\gg 1$), as captured by the indicator $S$, while
the NILC map has the lowest level.

Regarding the deviation from Gaussianity as detected by the kurtosis
indicator $K$, Table~\ref{Chi2-Table} shows again that the HILC presents the
largest deviation followed by the ILC and NILC. To the extent that
$\chi^2/\text{dof}\,$ is considerably greater than one, all these
full-sky foreground-reduced maps also present a significant deviation
from Gaussianity as captured here by the kurtosis indicator.

The above results of our statistical analysis given in Fig.~\ref{Fig3}
and gathered together in Table~\ref{Chi2-Table} (and also supported by
Tables~\ref{Skew-dev_prob_full} and~\ref{Skew-dev_prob_masked} of
Appendix~\ref{App_A}) show a significant
deviation from  Gaussianity in five-year full-sky foreground-reduced
(ILC, NILC and HILC) maps as detected by both
the skewness and the kurtosis indicators $S$ and $K$.
A pertinent question that arises here is how this
analysis of Gaussianity for the full-sky foreground-reduced maps is
modified if one uses the \emph{KQ75} mask, which was recommended
by the WMAP team for tests of Gaussianity of the five-year band maps.
Furthermore, the combination of the full-sky and mask analyses
should provide information on the reliability  of the foreground-reduced
maps as appropriate reconstructions of the full-sky CMB.

Figure~\ref{Fig4} shows the power spectra $|S_{\ell}
- \overline{S}_{\ell}|$ (left) and $|K_{\ell} - \overline{K}_{\ell}|$ (right)
calculated from five-year foreground-reduced \emph{KQ75} masked maps.
This figure along with Fig.~\ref{Fig3} show a significant reduction in the
level of deviation from Gaussianity when the foreground-reduced ILC, HILC,
and NILC maps are masked. To quantify this reduction we have recalculated
$\chi^2/\text{dof}\,$  for these input maps with the \emph{KQ75} mask, and have
collected the results in Table~\ref{Chi2-Table-masked}. The comparison
of Table~\ref{Chi2-Table} and Table~\ref{Chi2-Table-masked} shows
quantitatively the reduction of the level of Gaussianity for the
case of CMB masked maps.%
\footnote{Incidentally, this reduction is also revealed through (and agrees
with) the comparison of Table~\ref{Skew-dev_prob_full} with
Table~\ref{Skew-dev_prob_masked}, and of Table~\ref{Kurt-dev_prob_full}
with Table~\ref{Kurt-dev_prob_masked}.}

In the above analyses we have followed the five-year WMAP recommendation
for tests of Gaussianity and thus used the mask the \emph{KQ75}, which is slightly
more conservative than the  \emph{Kp0} (the\emph{KQ75} sky cut is
$28.4\%$ while the \emph{Kp0} cut is $24.5\%\,$). A pertinent question at
this point is how the above results are modified if the less
conservative \emph{Kp0} mask is used. We have examined this issue
by calculating the power spectra $S_\ell$ and $K_\ell$ and the
$\chi^2/\text{dof}\,$ from $S$ and $K$ maps obtained from the
ILC, NILC, and HILC input maps with the \emph{Kp0} mask. The result
of this analysis is given in Table~\ref{Chi2-Table-kp0mask}.

A comparison between Tables~\ref{Chi2-Table-masked} and~\ref{Chi2-Table-kp0mask}
shows that in general the value of $\chi^2/\text{dof}\,$ increases for
both indicators when the less conservative mask  \emph{Kp0} is used.
We note that the changes in $\chi^2/\text{dof}\,$ values are greater
for the HILC, though.

The comparison between Fig.~\ref{Fig3}  and Fig.~\ref{Fig4}, and
Tables~\ref{Chi2-Table} and Table~\ref{Chi2-Table-masked} along with
the tables of the Appendix~\ref{App_A} clearly provides quantitative information
on the suitability of the foreground-reduced maps as Gaussian
reconstructions of the full-sky CMB, and makes apparent the
relevant role of the mask \emph{KQ75} in reducing significantly
the level of non-Gaussianity in these foreground-reduced maps.

The calculations of our non-Gaussianity indicators require the
specification of some quantities whose choice could in principle
affect the outcome of our calculations.
To test the robustness of our scheme, hence of our results, we
studied the effects of changing in the parameters employed in
the calculation of our indicators.
We found that the $S$ and $K$ angular power spectra do not
change appreciably as we change the resolution of CMB temperature
maps used 
and the number of point-centers of the caps with values $768$,
$3\,072$ and $12\,288$ (see Ref.~\cite{Bernui-Reboucas2009} for
more details on the robustness of this method).

Concerning the robustness of the above analyses with the \emph{KQ75} mask
some additional words of clarification are in order here. First, we note
that the calculations of the $S-$maps and $K-$maps by scanning the CMB
masked maps sometimes include caps whose center is within or close to
the \emph{KQ75} masked region.
In these cases, the calculations of the $S$ and $K$ indicators are made with
a smaller number of pixels, which clearly introduce additional statistical noise
as compared to the full-sky map cases.
In order to minimize this effect we have scanned the CMB masked sky with
spherical caps of aperture  $\gamma = 90^{\circ}$, and for the sake of
uniformity we have used caps with the same aperture for the full-sky maps.
We note, however, that full-sky foreground-reduced analysis does not
change significantly if one uses smaller apertures as, for example,
$\gamma \simeq 60^{\circ}$.

\section{Concluding remarks}

The detection or nondetection of primordial non-Gaussianity in
the CMB data is essential to discriminate or even exclude classes of
inflationary models. It can also be used to test alternative
scenarios of the primordial universe.
There are, however, several non-primordial effects that can also
produce  non-Gaussianity. This makes the extraction of
a possible primordial non-Gaussianity a rather difficult endeavor.
Since different indicators can in principle
provide information about distinct forms of non-Gaussianity,
it is important to test CMB data for non-Gaussianity by
using different estimators to quantify and/or constrain
its amount in order to extract information about their possible
sources.

Most of the Gaussianity analyses of CMB data have been performed
with frequency band maps. In these studies, to deal with the galactic
diffuse foreground emission, masks have been employed.
However, 
a full-sky foreground-reduced map seems to be potentially more appropriate
to test for Gaussianity the CMB data.%
\footnote{In reality, the full-sky map seems to be the most suitable for a
number of other issues, including the test of statistical
isotropy, the search for evidence of a North-South asymmetry in CMB data,
and signatures of a possible nontrivial cosmic topology, for example.}
The five-year version of the ILC map has been suggested as a full-sky
map suitable for large angular scales analyses~\cite{ILC-3yr-Hishaw},
even though the WMAP team has not performed a battery of non-Gaussianity
tests on this map~\cite{ILC-5yr-Hishaw}.

In this paper we have performed an analysis of Gaussianity of the
available five-year full-sky foreground-reduced  maps.
To this end, we have used two  new non-Gaussianity indicators  based
on skewness and kurtosis of large-angle patches of CMB maps,
which provide a measure of departure from Gaussianity on large
angular scales~\cite{Bernui-Reboucas2009}.
We have shown that the full-sky five-year foreground-reduced maps
(ILC, HILC and NILC) present a significant deviation from Gaussianity,
which varies with the foreground-reducing procedures.
We have  established which of these full-sky foreground-reduced
maps exhibit the highest and the lowest level of non-Gaussianity.

We have also masked the foreground-reduced maps with  \emph{KQ75} and \emph{Kp0}
masks and performed a quantitative analysis of deviation from Gaussianity of these
maps. The comparison of the full-sky and masked analyses (see Fig.~\ref{Fig3}
and Fig.~\ref{Fig4}; and Tables~\ref{Chi2-Table}, \ref{Chi2-Table-masked}
and~\ref{Chi2-Table-kp0mask})
shows a significant reduction in the levels of non-Gaussianity when the
masks are employed,  which in turn provides indications on the suitability of the
foreground-reduced maps as Gaussian reconstructions of the full-sky CMB.

Finally, when we were in the process of rewriting a revised version of this
paper, by taking into account the referee's recommendations, the seven-year WMAP
CMB data were released, including a new version of the full-sky
foreground-reduced ILC map~\cite{ILC-7yr-Gold}. We have considered this latest
foreground-reduced ILC  map, and performed a complete additional
analysis of the Gaussianity of the five and seven-year versions of the ILC maps,
whose details are given in Apenddix~\ref{App_B}.%
\footnote{Note that there is no available seven-year HILC and NILC maps.}
The main result of this appendix is that the full-sky seven-year
foreground-reduced ILC map also present a
significant deviation from Gaussianity, which again is reduced substantially
when the \emph{KQ75} mask is employed. In this way, our results are robust
with respect to seven-year WMAP CMB data.

\vspace{-0.5cm}
\begin{acknowledgments}
This work is supported by Conselho Nacional de Desenvolvimento Cient\'{i}fico e
Tecnol\'ogico (CNPq) -- Brasil, under Grant No.\ 472436/2007-4.
M.J.R. and A.B. thank CNPq  for the grants under which this work was carried out.
We are also grateful to A.F.F. Teixeira for reading the manuscript
and indicating the omissions and misprints.
We acknowledge the use of the Legacy Archive for Microwave Background Data
Analysis (LAMBDA).
Some of the results in this paper were derived using the HEALPix
package~\cite{Gorski-et-al-2005}.
\end{acknowledgments}

\appendix
\section{}  \label{App_A}

Clearly from $1\,000$ MC maps one can calculate for each $\ell$ one thousand
values of both $S^{\,\mathbf{i} }_{\,\ell}$ and  $K^{\,\mathbf{i} }_{\,\ell}$
($\,\mathbf{i}= 1,\,\,\cdots,1\,000\,$) and the corresponding mean values
$\overline{S}_{\ell}$ and $\overline{K}_{\ell}$.
For the sake of brevity in what follows we focus on the skewness
indicator $S$, but completely similar calculations were used to have
the probabilities for kurtosis indicator $K$.

\begin{table}[t]
\begin{center}
\begin{tabular}{cccc} 
\hline  \hline 
$\ell$  & \ \ NILC [full-sky] \ \ & \ \ HILC [full-sky] \ \ & \ \ ILC [full-sky]  \\
\hline  
$1$ & \ \ \ $51.7\%$    & \ \ \ ${\cal O}(10^{-3})\%$  & \ \ \ $3.7\%$     \\
$2$ & \ \ \ ${\cal O}(10^{-3})\%$ & \ \ \ ${\cal O}(10^{-3})\%$  & \ \ \ ${\cal O}(10^{-3})\%$  \\
$3$ & \ \ \ $0.1\%$      & \ \ \ $0.1\%$      & \ \ \ $0.4\%$     \\
$4$ & \ \ \ ${\cal O}(10^{-3})\%$  & \ \ \ ${\cal O}(10^{-3})\%$ & \ \ \ $0.1\%$     \\
$5$ & \ \ \ ${\cal O}(10^{-3})\%$  & \ \ \ ${\cal O}(10^{-3})\%$ & \ \ \ $0.1\%$     \\
$6$ & \ \ \ ${\cal O}(10^{-3})\%$  & \ \ \ ${\cal O}(10^{-3})\%$  & \ \ \ \%${\cal O}(10^{-3})$  \\
$7$ & \ \ \ $1.2\%$      & \ \ \ $0.1\%$       & \ \ \ $1.3\%$     \\
$8$ & \ \ \ $0.1\%$      & \ \ \  $0.1\%$      & \ \ \ $0.1\%$     \\
$9$ & \ \ \ ${\cal O}(10^{-3})\%$  & \ \ \ ${\cal O}(10^{-3})\%$  & \ \ \ ${\cal O}(10^{-3})$  \\
$10$&\ \ \ $0.1\%$      & \ \ \ ${\cal O}(10^{-3})\%$   & \ \ \ $0.1\%$     \\
\hline \hline
\end{tabular}
\end{center}
\caption{The probability (percentage) of occurrence of  the multipole values
$S_{\ell}$ calculated from the data in the set \{$S^{\,\mathbf{i} }_{\ell}$\}
of values computed from  MC Gaussian CMB maps. The data
from the five-year full-sky foreground-reduced NILC, HILC and ILC
maps were used.}
\label{Skew-dev_prob_full}
\end{table}

\begin{table}[t]
\begin{center}
\begin{tabular}{cccc} 
\hline \hline 
$\ell$  & \ \  NILC [\emph{KQ75}] \ \ & \ \ HILC [\emph{KQ75}] \ \ & \ \ ILC [\emph{KQ75}] \\
\hline  
$1$ & \ \ \ $76.7\%$ & \ \ \ $34.1\%$ & \ \ \  $73.7\%$ \\
$2$ & \ \ \ $69.1\%$ & \ \ \ $12.8\%$ & \ \ \  $42.1\%$ \\
$3$ & \ \ \ $4.6\%$  & \ \ \ $3.7\%$    & \ \ \  $5.7\%$ \\
$4$ & \ \ \ $7.2\%$  & \ \ \ $8.9\%$    & \ \ \  $16.3\%$ \\
$5$ & \ \ \ $2.7\%$  & \ \ \ $2.2\%$    & \ \ \  $11.3\%$ \\
$6$ & \ \ \ $2.4\%$  & \ \ \ $1.5\%$    & \ \ \  $4.0\%$ \\
$7$ & \ \ \ $51.4\%$ & \ \ \ $57.6\%$ & \ \ \  $54.1\%$ \\
$8$ & \ \ \ $4.9\%$  & \ \ \ $10.2\%$  & \ \ \  $40.6\%$ \\
$9$ & \ \ \ $7.0\%$  & \ \ \ $13.8\%$  & \ \ \  $55.0\%$ \\
$10$& \ \ \ $3.9\%$ & \ \ \ $8.4\%$    & \ \ \  $27.7\%$ \\
\hline \hline
\end{tabular}
\end{center}
\caption{The probability (percentage) of occurrence of 
multipole values $S_{\ell}$ calculated from the data in the set \{$S^{\,\mathbf{i} }_{\ell}$\}
of values computed from MC Gaussian CMB maps. The data from the
five-year NILC, HILC and ILC \emph{KQ75} masked maps were used.}
\label{Skew-dev_prob_masked}
\end{table}

\begin{table}[t] 
\begin{center}
\begin{tabular}{cccc} 
\hline \hline 
$\ell$  & \ \ NILC [full-sky] \ \ & \ \ HILC [full-sky] \ \ & \ \ ILC [full-sky] \\
\hline  
$1$ & \ \ \ $0.6\%$     & \ \ \ ${\cal O}(10^{-3})\%$  & \ \ \ ${\cal O}(10^{-3})\%$ \\
$2$ & \ \ \ ${\cal O}(10^{-3})\%$ & \ \ \ ${\cal O}(10^{-3})\%$  & \ \ \ ${\cal O}(10^{-3})\%$ \\
$3$ & \ \ \ $0.1\%$     & \ \ \ ${\cal O}(10^{-3})\%$  & \ \ \ $0.1\%$      \\
$4$ & \ \ \ ${\cal O}(10^{-3})\%$ & \ \ \ ${\cal O}(10^{-3})\%$  & \ \ \ ${\cal O}(10^{-3})\%$ \\
$5$ & \ \ \ $0.1\%$   & \ \ \ ${\cal O}(10^{-3})\%$  & \ \ \ ${\cal O}(10^{-3})\%$   \\
$6$ & \ \ \ $0.1\%$   & \ \ \ ${\cal O}(10^{-3})\%$  & \ \ \ $0.1\%$      \\
$7$ & \ \ \ $0.4\%$   & \ \ \ ${\cal O}(10^{-3})\%$  & \ \ \ ${\cal O}(10^{-3})\%$ \\
$8$ & \ \ \ $0.1\%$   & \ \ \ ${\cal O}(10^{-3})\%$  & \ \ \ ${\cal O}(10^{-3})\%$ \\
$9$ & \ \ \ $0.1\%$   & \ \ \ ${\cal O}(10^{-3})\%$  & \ \ \ ${\cal O}(10^{-3})\%$ \\
$10$&\ \ \ $0.1\%$   & \ \ \ ${\cal O}(10^{-3})\%$  & \ \ \ ${\cal O}(10^{-3})\%$ \\
\hline \hline
\end{tabular}
\end{center}
\caption{The probability (percentage) of occurrence of the  
multipole values $K_\ell$ calculated from the data in the set \{$K^{\,\mathbf{i} }_{\ell}$\}
of values computed from MC Gaussian CMB maps. The data
from the five-year full-sky foreground-reduced NILC, HILC and ILC
maps were utilized.}
\label{Kurt-dev_prob_full}
\end{table}

\begin{table}[!h]
\begin{center}
\begin{tabular}{cccc} 
\hline \hline 
$\ell$  & \ \  NILC [\emph{KQ75}] \ \ & \ \ HILC [\emph{KQ75}] \ \ & \ \ ILC [\emph{KQ75}] \\
\hline  
$1$ & \ \ \ $87.0\%$ & \ \ \ $45.1\%$  & \ \ \  $81.1\%$ \\
$2$ & \ \ \ $2.3\%$   & \ \ \ $1.9\%$    & \ \ \  $23.6\%$ \\
$3$ & \ \ \ $85.6\%$ & \ \ \ $29.3\%$  & \ \ \  $89.1\%$ \\
$4$ & \ \ \ $2.3\%$   & \ \ \ $9.0\%$   & \ \ \  $57.1\%$  \\
$5$ & \ \ \ $1.3\%$   & \ \ \ $1.8\%$   & \ \ \  $4.7\%$    \\
$6$ & \ \ \ $82.6\%$  & \ \ \ $4.2\%$  & \ \ \  $98.1\%$  \\
$7$ & \ \ \ $8.8\%$    & \ \ \ $1.7\%$  & \ \ \  $46.9\%$  \\
$8$ & \ \ \ $96.6\%$  & \ \ \ $93.2\%$ & \ \ \  $86.6\%$ \\
$9$ & \ \ \ $0.9\%$    & \ \ \ $0.8\%$   & \ \ \  $5.0\%$   \\
$10$& \ \ \ $65.0\%$ & \ \ \ $54.2\%$ & \ \ \  $97.7\%$ \\
\hline \hline
\end{tabular}
\end{center}
\caption{The probability (percentage) of occurrence of the 
multipole value $K_\ell$ calculated from the data in the set \{$K^{\,\mathbf{i} }_{\ell}$\}
of values computed from MC Gaussian CMB maps. The data
from the five-year foreground-reduced NILC, HILC and ILC \emph{KQ75} masked
maps were employed.}
\label{Kurt-dev_prob_masked}
\end{table}

With the MC values $S^{\,\mathbf{i} }_{\,\ell}$ and the mean $\overline{S}_{\ell}$
one can calculate the percentages of values of the deviations
$|S^{\,\mathbf{i} }_{\ell} - \overline{S}_{\ell}|$ calculated from $1\,000\,$ MC Gaussian
maps which are smaller than $|S_{\ell} - \overline{S}_{\ell}|$ with $S_{\ell}$ obtained
from the data (full-sky and masked maps). For each multipole
this number indicates how unlikely are the occurrences of the values
obtained from the data (input maps) for that multipole in
the set of values  $\{S^{\,\mathbf{i} }_{\ell}\}$ obtained from MC Gaussian maps.
In this way one can calculate the probability of occurrence of a given multipole
value $S_{\ell}$ (obtained from the data) in the set of MC values
(obtained from the MC maps) for each foreground-reduced
CMB map. 
In Tables~\ref{Skew-dev_prob_full}, \ref{Skew-dev_prob_masked},
\ref{Kurt-dev_prob_full} and \ref{Kurt-dev_prob_masked} we collect
together the results of such calculations.

\begin{figure*}[!t] 
\begin{center}
\includegraphics[width=8.8cm,height=5.6cm]{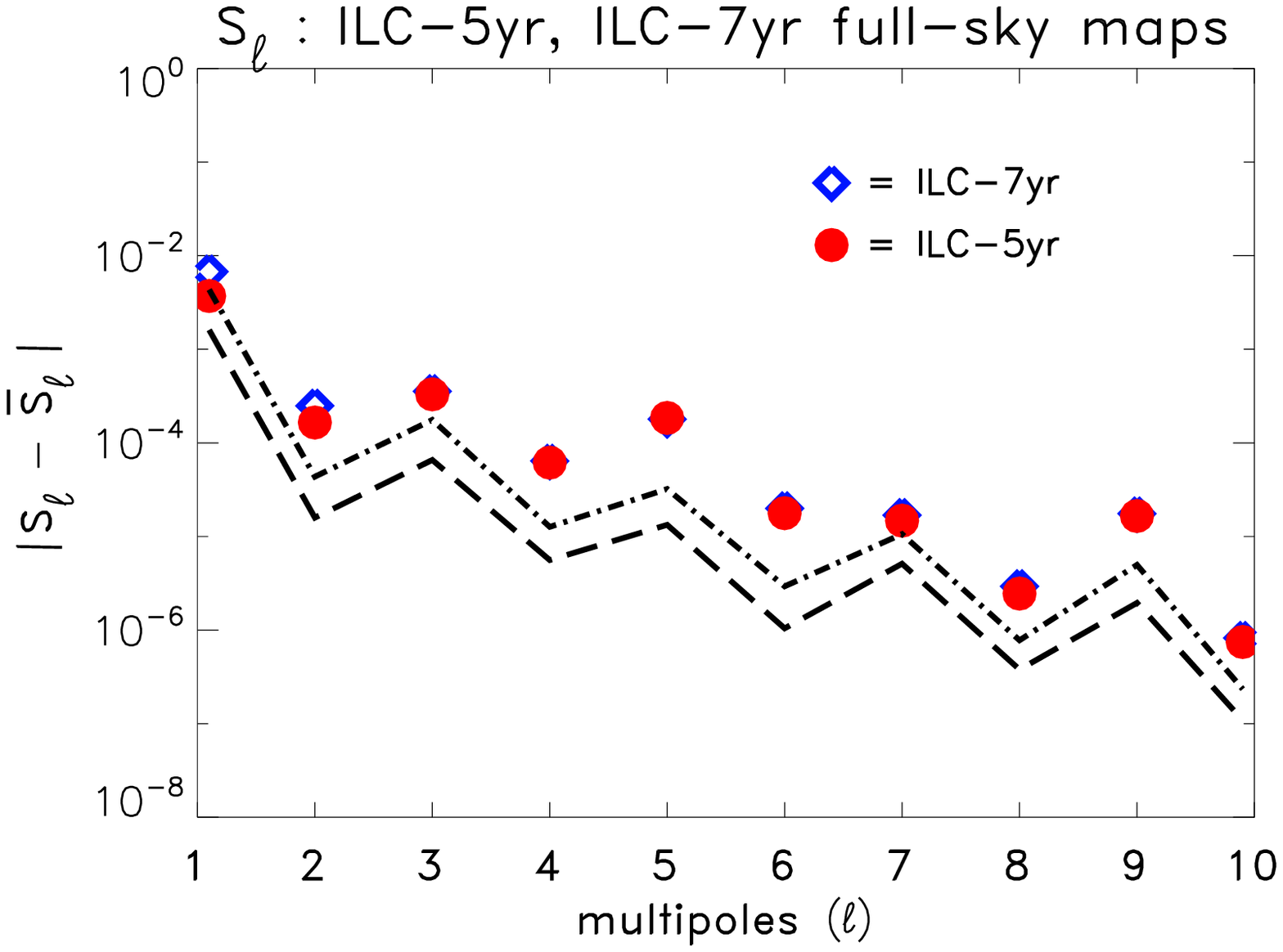}
\includegraphics[width=8.8cm,height=5.6cm]{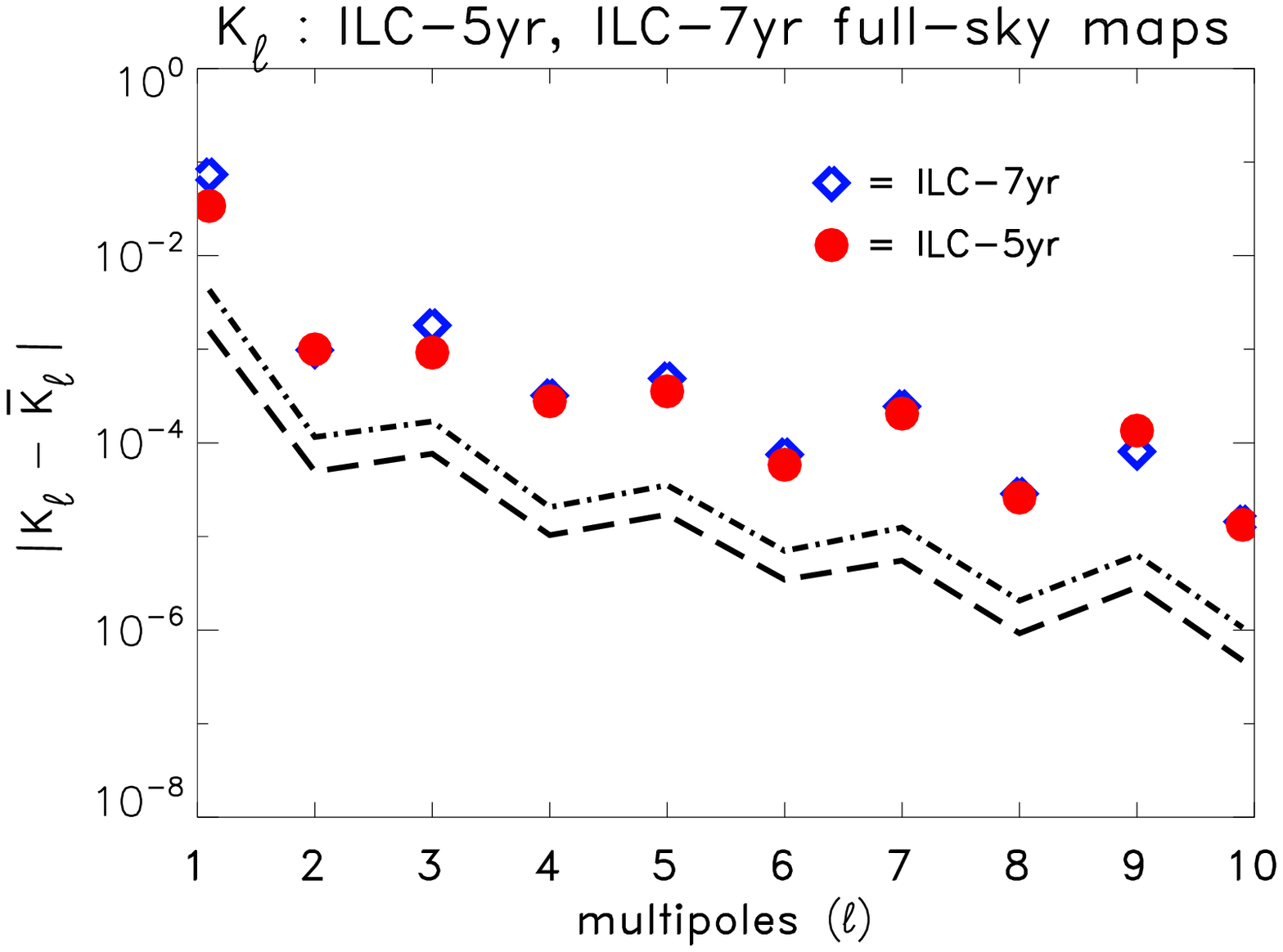}
\caption{Differential power spectrum of skewness $|S_{\ell} - \overline{S}_{\ell}|$
(left) and kurtosis $|K_{\ell} - \overline{K}_{\ell}|$ (right) indicators calculated
from the full-sky foreground-reduced ILC input map obtained from
the WMAP five-year and seven-year data. The $68\%$ and $95\%$ confidence levels are
indicated, respectively, by the dashed and dash-dotted lines.}
\label{Fig5}
\end{center}
\end{figure*}

\begin{table*}[!t]
\begin{center}
\begin{tabular}{ccc} 
\hline \hline 
\ \  & \ \ \ $\chi^2_{}$ for $S_\ell$ [full-sky] \ \ \ & \ \ \ $\chi^2_{}$ for $K_\ell$ [full-sky]
\ \ \      \\
\hline
ILC5               \ \  & \ \ \ $35.7$      \ \ \  & \ \ \  $2368$     \ \ \     \\
ILC7               \ \  & \ \ \ $103$      \ \ \   & \ \ \  $10507$   \ \ \      \\
\hline \hline
\end{tabular}
\end{center}
\caption{Results of the $\chi^2$ test  to determine
the goodness of fit for $S_{\ell}$ and $K_{\ell}$ multipole values calculated
from the foreground-reduced ILC5 and ILC7 full-sky maps as compared to
the expected multipoles values from the  MC Gaussian full-sky maps.}
\label{Chi2-Table-ILC5-7_full-sky}
\end{table*}

Thus, for example, from Table~\ref{Skew-dev_prob_full}
we have for the full-sky NILC, HILC, and ILC maps, respectively,
the probability of occurrence of the $S_{6}$ values
(in the set of MC values) is ${\cal O}(10^{-3})\%$, whereas from Table~\ref{Kurt-dev_prob_full}
the probability for $K_6$ is, respectively,  $0.1\%$, ${\cal O}(10^{-3})\%$
and $0.1\%$ for  the full-sky NILC, HILC, and ILC input maps.

The comparison of Table~\ref{Skew-dev_prob_full} with
Table~\ref{Skew-dev_prob_masked}, and  of
Table~\ref{Kurt-dev_prob_full} with Table~\ref{Kurt-dev_prob_masked}
shows that the role of the \emph{KQ75} mask is to cut down significantly the
level of deviation from Gaussianity for all multipoles $S_\ell$
and $K_\ell$ obtained from the foreground-reduced input maps.
This is clear because the probabilities  of occurrences for these multipoles
values in the set of MC multipole values increase substantially when the
mask is employed.

Although the estimates of probabilities collected in these tables
give a clear quantitative indication of deviation from Gaussianity
an overall assessment of the power spectra $S_\ell$ and $K_\ell$
can be obtained through $\chi^2$ test of the goodness of fit
for $S_{\ell}$ and $K_{\ell}$ from the data as compared to
the expected multipoles values obtained from the Gaussian
MC maps. This point is discussed in Section~\ref{Anal_Resul}.

\section{}  \label{App_B}

While we were in the final phase of writing a modified version of this paper
a new version of the full-sky foreground-reduced ILC map was released
by the WMAP team~\cite{ILC-7yr-Gold}. Since there is no available version
of the NILC and HILC maps obtained from the seven-year WMAP data to
be considered, here we present the results of a comparative analysis of
deviation from Gaussianity performed by using the five and seven-year
versions of the ILC as input maps. As the calculations are similar to
those of Section~\ref{Anal_Resul} we refer the readers to that section
for more details.

Figure~\ref{Fig5} shows the differential power spectra calculated from
the full-sky five and seven-year foreground-reduced ILC input maps
(ILC5 and ILC7, for short). Apart from some local deviation of the
deviations $|S_{\ell} - \overline{S}_{\ell}|$ and
$|K_{\ell} - \overline{K}_{\ell}|$ this figure shows a deviation
from Gaussianity, which is quantified in
Table~\ref{Chi2-Table-ILC5-7_full-sky}.

\begin{figure*}[!t]
\begin{center}
\includegraphics[width=8.8cm,height=5.6cm]{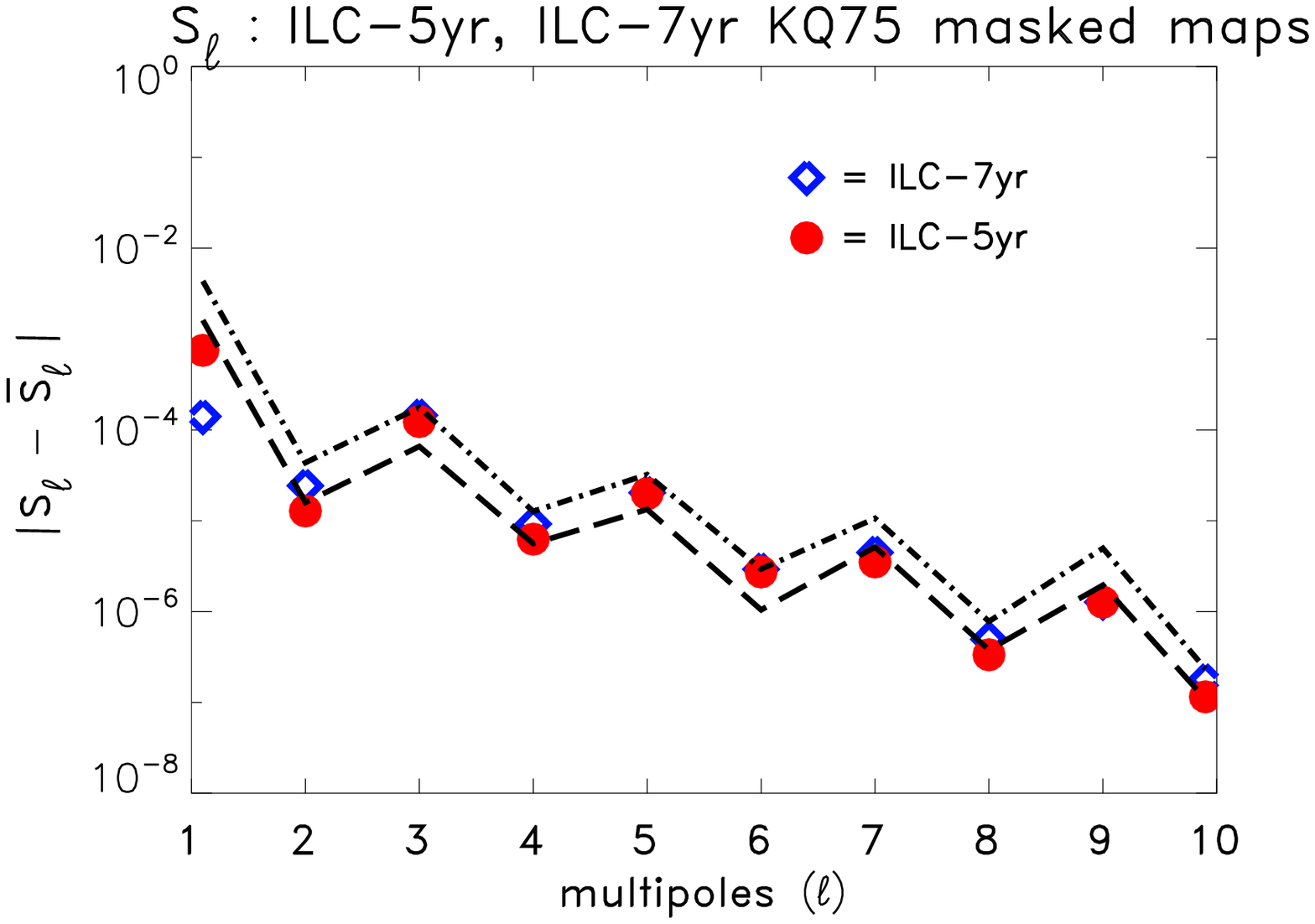}
\includegraphics[width=8.8cm,height=5.6cm]{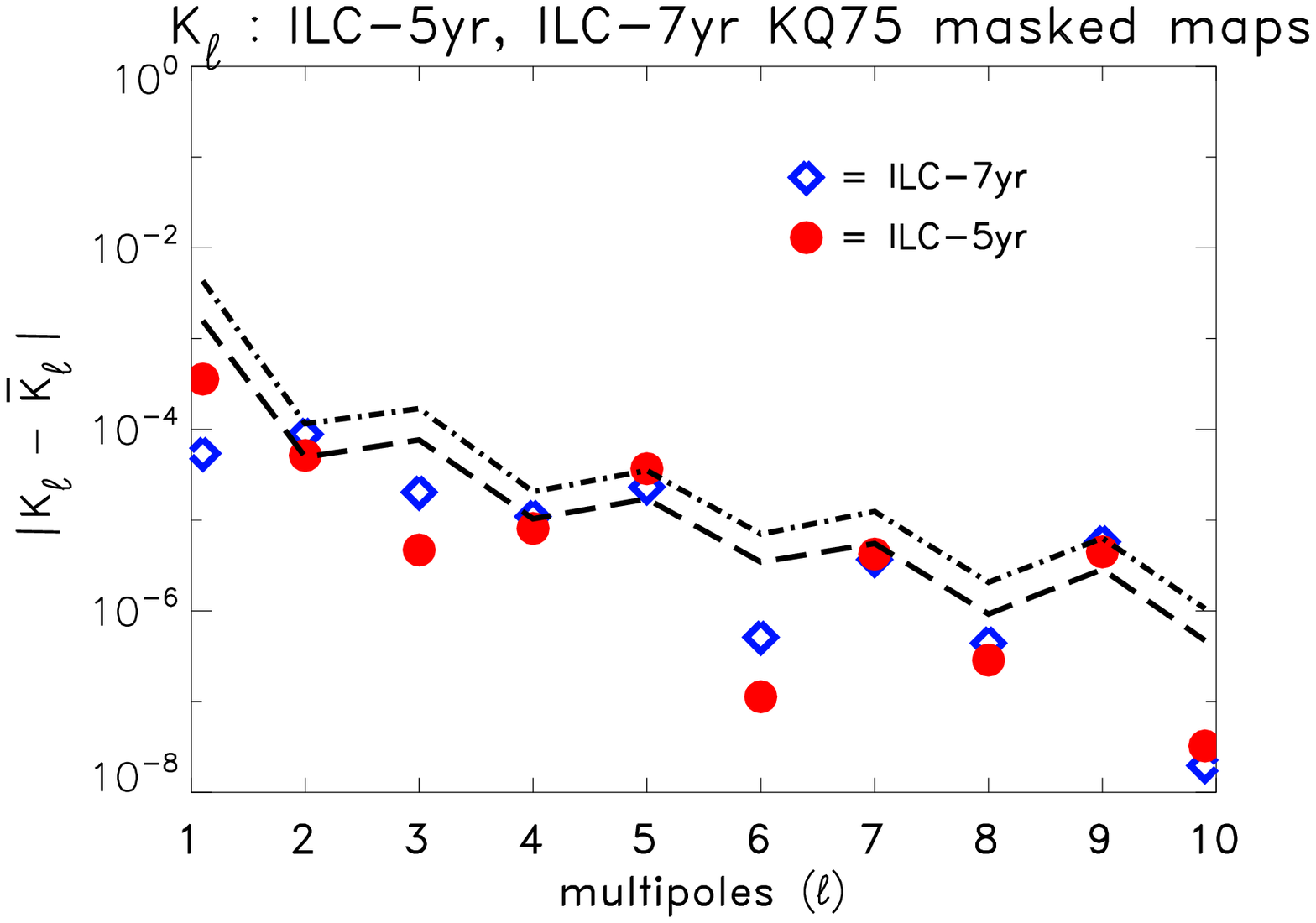}
\caption{Differential power spectrum of skewness
$|S_{\ell} - \overline{S}_{\ell}|$ (left) and  kurtosis $|K_{\ell} - \overline{K}_{\ell}|$
(right) indicators calculated from the five-year and seven-year
foreground-reduced ILC input maps with a \emph{KQ75} mask.
The $68\%$ and $95\%$ confidence levels are
indicated, respectively, by the dashed and dash-dotted lines.}
\label{Fig6}
\end{center}
\end{figure*}

\begin{table*}[!t]
\begin{center}
\begin{tabular}{ccc} 
\hline \hline 
\ \  & \ \ \ $\chi^2_{}$ for $S_\ell$ [\emph{KQ75}] \ \ \ & \ \ \ $\chi^2_{}$ for $K_\ell$ [\emph{KQ75}]
\ \ \      \\
\hline
ILC5               \ \  & \ \ \ $1.2$      \ \ \  & \ \ \ $0.4$     \ \ \     \\
ILC7               \ \  & \ \ \ $0.7$      \ \ \ & \ \ \  $0.2$   \ \ \      \\
\hline \hline
\end{tabular}
\end{center}
\caption{Results of the $\chi^2$ test  to determine
the goodness of fit for $S_{\ell}$ and $K_{\ell}$ multipole values
calculated from the foreground-reduced ILC5 and ILC7 maps
with a \textit{KQ75} mask as compared to the expected
multipoles values from the Gaussian MC masked maps.}
\label{Chi2-Table-ILC5-7_KQ75}
\end{table*}

It is interesting to note that the deviation from Gaussianity as
measured by our indicators is greater for the ILC7 than for the ILC5 input map.
Concerning this point some words of clarification are in order here.
First, we note that the details of the algorithm used to
compute the ILC7 maps are the same as those of the ILC5 map.
However, to take into account the most recent updates to the calibration
and beams, the frequency weights for each of the 12 regions (in which
the sky is subdivided in the ILC method) are slightly different in
the calculation of the ILC7 map.
Second, the difference between the ILC7 and ILC5 maps is a map
whose small-scale differences are consistent with the pixel noise,
but with a large-scale dipolar component, with the large-scale
differences being consistent with a change in dipole of
6.7 $\mu$K\cite{ILC-7yr-Gold}. Thus, the resultant ILC7 map
is not indistinguishable from the ILC5 map, and the differences
between them have been captured by our indicators.

Figure~\ref{Fig6} shows the differential power spectra calculated from
a five-year and seven year version of the foreground-reduced ILC maps
with a \emph{KQ75} mask. This figure along with Fig.~\ref{Fig5} show
a significant reduction in the level of deviation from Gaussianity
when both ILC5 and ILC7 are masked.
To quantify this reduction we have calculated
$\chi^2/\text{dof}\,$  for these input maps with the \emph{KQ75} mask, and have
collected the results in Table~\ref{Chi2-Table-ILC5-7_KQ75}. The comparison
of Table~\ref{Chi2-Table-ILC5-7_full-sky} and Table~\ref{Chi2-Table-ILC5-7_KQ75}
shows quantitatively the reduction of the level of Gaussianity for the case of
CMB masked maps.


\end{document}